\documentstyle{article}
\def\PLA{{\em Phys. Lett.}  A}
\def\NCB{{\em Il Nuovo Cimento} B}
\def\LNC{{\em Lett. N. Cimento}}
\def\JMP{{\em J.Math.Phys.}}
\def\IM{{\em Indag. Math.} A}

\def\PLB{{\em Phys. Lett.}  B}
\def\PRL{{\em Phys. Rev. Lett.}}

\def\AM{{\em Acta Math.}}
\def\CMP{{\em Commun. Math. Phys.}}
\def\be{\begin{equation}}
\def\ee{\end{equation}}
\def\bea{\begin{eqnarray}}
\def\eea{\end{eqnarray}}

\vspace{.5cm}
\begin{document}
\begin{center}
\vspace{1.5cm}
{\LARGE\bf Liouville Integrability\\
 of the Schr\"odinger Equation} 
\footnote{Supported in part by the italian Ministero
dell' Universit\`a e della Ricerca Scientifica e Tecnologica.}
\footnote{Talk given at Proceedings of Caserta Conference 
{\it Mesoscopic Systems and Quantum-like Phenomena} (1996).}
\vspace{1.5cm}
\par
by
\par
{\bf Gaetano Vilasi}
\medskip
\par
{\it Dipartimento di Fisica Teorica e smsa, 
Universit\`a di Salerno,\\
Via S. Allende, I-84081 Baronissi (SA), Italy.~(vilasi@salerno.infn.it)}
\par
{\it Istituto Nazionale di Fisica Nucleare, Sezione di Napoli, Italy.}
medskip
\centerline{PACS Nos.:03.20+i,03.65-W}
\end{center}
\vspace{3.cm}
\begin{abstract}
Canonical coordinates for both the Schr\"odinger 
and the nonlinear Schr\"odinger equations are introduced,
making more trans\-parent their Ha\-mil\-tonian structures. It is shown that 
the  Schr\"odinger equation, considered as a classical field theory, 
shares with the nonlinear Schr\"odinger, and more generally with  
Liouville completely integrable field theories, the
existence of a {\sl recursion operator} which allows for the 
construction of infinitely 
many conserved functionals pairwise commuting with respect to the 
corresponding Poisson bracket. 
The approach may provide a good starting point to get a clear 
interpretation of Quantum Mechanics in the general setting, provided 
by Stone-von Neumann theorem, of Symplectic Mechanics. 
It may give new tools to solve in the general case the inverse 
problem of Quantum Mechanics.

\end{abstract} 

\vfill\eject

\section*{Introduction}

Last two decades have shown the exciting prospects of tackling
nonlinear field theories in two space time dimensions nonperturbatively by 
exploiting their complete integrability properties \cite{FT74,DMN76}.
\par
Relevant progresses in the analysis of these systems 
were the introduction of the Lax 
Representation \cite{La68}, the Zakharov-Shabat scheme \cite{JZ79}  
and AKNS method \cite{Ka76,CD82}.
\par
Lax Representation played an important role in formulating the inverse 
scattering method \cite{NMPZ84} which allows
for the solution of the Cauchy problem by means of the Gel'fand-Levitan 
-Marchenko formula \cite{GL51,Marc55}.
\par
A universal feature of almost such systems is that they are Hamiltonian 
systems with infinitely many  degrees of freedom \cite{GGKM67}. 
The inverse scattering method was then read 
as a canonical transformation from generic coordinates (potentials) to 
action-angle variables \cite{FZ71}.
\par
This fact made it only natural a formulation of 
{\it a priori} criteria of 
integrability,  by methods more 
directly related to Group Theory \cite{Sc87,NMPZ84} 
and to familiar 
procedures of Classical Mechanics, looking at such systems as 
dynamics  on (infinite-dimensional) phase manifold 
\cite{Ol77,Ma78,Ma80}$^{\!,\,}$\cite{Vi80}$^{\!-\,}$\cite{ZK84,Bo96}.
\par	
This point of view was also suggested by the occurrence in such 
models of a peculiar operator , the so called 
{\it recursion operator} \cite{La68}, relevant for the effectiveness 
of the method, which naturally fits in this geometrical setting as 
a mixed tensor field on the phase manifold $M$.
\par
In terms of such an operator the classical Liouville theorem on the 
integrability can be extended also to the infinite dimensional case. 
The same operator can be used to deal with Burgers equation \cite{DMSV83}.
\par
It will be shown that, in complete analogy with the
case of the nonlinear Schr\"odinger equation, such an invariant 
tensor field exists for the Schr\"odinger equation too.
\par
Some years ago it was suggested \cite{St66} the use of complex canonical 
coordinates in the formulation of a  generalised dynamics including
classical and quantum mechanics as special cases. In the same spirit
a somehow dual viewpoint is proposed : to formulate  Quantum Mechanics
in terms of {\it realified} vector spaces.
\par
By using the Stone-von Neumann theorem a quantum mechanical system is 
associated with a vector field on some Hilbert space ({\sl Schr\"odinger
picture}) or a vector field, i.e. a derivation, on the algebra of 
observables ({\sl Heisemberg picture}).
\par
In Classical Mechanics the analogous infinitesimal generator of 
canonical transformations is a vector field on a symplectic 
manifold (the {\sl phase space}).
\par
In order to use the familiar procedures of Classical mechanics, 
we need to real off $L_2(Q, {\bf C})$, the Hilbert space  of 
square integrable complex functions defined on the configuration space $Q$, 
as a symplectic manifold or, more specifically, as a cotangent bundle. 
We shall see that it can be considered as $T^*(L_2(Q, {\bf R}))$,
~~$L_2(Q, {\bf R})$ denoting the Hilbert space of square integrable
real functions defined on $Q$.
\par
The approach is different from previous ones 
\cite{Ku85,Bl86,CP90,CDL96} also 
dealing with the integrability of quantum mechanical system in
the Heisemberg and Schr\"odinger picture.
\par
In order to make more transparent the geometrical and the physical 
content of the paper difficult technical aspects, which are 
however important in the context of infinite dimensional manifold 
as, for instance, the distinction \cite{CM74} between {\sl weakly} 
and {\sl strongly} not degenerate bilinear forms, or the inverse 
of a Schr\"odinger operator and so on,  will  not be addressed. 
We shall limit ourselves to observe that no serious difficulties 
arise working on an infinite dimensional manifold whose local model 
is a Banach space, as in that case the {\sl implicit function 
theorem} still holds true.
First section deals with Liouville integrability, the second one
with the Scr\"odinger equation and last with the nonlinear one. 

\section{Complete Integrability and Recursion Operators} 
Complete integrability of Hamiltonian systems with finitely many 
degrees of freedom is exhaustively characterised by the Liouville-Arnold 
theorem \cite{Li897,Po899,Ar76}. 
An alternative characterisation which may apply 
also to systems with infinitely many degrees of freedom can be 
given as follows.
Let $M$ denote a smooth differentiable manifold, ${\cal X}(M)$ and 
$\Lambda(M)$ vector and covector fields on $M$.  
With any   $(1,1)$   tensor field   $T$   on  $ M$, two endomorphisms  
~~${\hat T} : {\cal X}(M)\rightarrow {\cal X}(M)$~~and~~ 
${\check T} : \Lambda(M)\rightarrow \Lambda(M) $~~are associated:
\begin{equation}	
T(\alpha, X)  = <\alpha,  {\hat T}X > = <{\check T}\alpha, X>,      
\end{equation}
with   $X$ and  $\alpha$ belonging to ${\cal X}(M)$  and  $\Lambda(M)$  
respectively.	The Nijenhuis  tensor  \cite{FN56,Ni87},  or {\it torsion},  
of $T$ is  the (1,2) tensor field defined by:
\begin{equation}		
			N_T(\alpha,X,Y) = <\alpha, H_T(X,Y)>                                                      
\end{equation}
with the vector  field  $ H_T(X, Y)$   given by:
\begin{equation}		
			H_T(X, Y)  = [\widehat{{\cal L}_{{\hat T}X}T} - {\hat T}
\widehat{{\cal L}_XT}]Y                                       
\end{equation}
${\cal L}_X$ denoting the Lie's derivative with respect to $X$.
\par
\noindent {\bf{ Integrability Criterion}}~\footnote{The vector 
field $\Delta$ is not supposed to be Hamiltonian. Its 
Hamiltonian structure is generated by the hypothesis of the 
bidimensionality of the eigenspaces of T and $d \lambda \ne 0$.}
\smallskip
\par
{\it A dynamical vector field $\Delta$ which admits an invariant 
mixed tensor field T, with vanishing Nijenhuis tensor $N_{T}$ and 
bidimensional eigenspaces, completely separates in 1-degree 
of freedom dynamics. The ones associated with those degrees of 
freedom whose corresponding eigenvalues $\lambda$ are not stationary, 
are integrable and Hamiltonian} \cite{DMSV82,DMSV84,DSV85,LMV94}.
\par
An idea of the proof is given observing that the bidimensionality 
of eigen\-spa\-ces
of $T$ and the condition $N_T=0$ imply the following form for $T$ 
\smallskip
\par
$T = \sum_{i} \lambda_{i} \biggl ({\delta\over \delta 
\lambda^{i}}\otimes \delta \lambda^{i} +{\delta\over \delta\phi^{i}} 
\otimes \delta \phi^{i} + {\delta\over \delta \phi^{i}} \otimes 
\delta \lambda^{i}\biggr ) + \sum^{2}_{\ell =1} 
\int_{0}^{k} dk~~ k {\delta\over \delta \psi^{\ell}_{k}}\otimes
\delta \psi^{\ell}(k)$
\smallskip
\par
The invariance of $T$ ~$ ({\cal L}_{\Delta} T = 0)$ implies for 
$\Delta$ the form
\par
$\Delta = \sum_{i=1}^{n} \Delta^{i}
(\lambda^{i}){\delta\over {\delta\phi^{i}}} + \sum^{2}_{\ell =1}  
\int dk \Delta^{\ell} (k)\biggl (\psi^{1}(k),\psi^{2}(k)\biggr )
{\delta\over {\delta\psi^{\ell}(k)}}$
\smallskip
\par
whose associated equations are:
\smallskip
\par
$\dot \psi^{1} (k) = \Delta^{1,k} (\psi^{1,(k)}, \psi^{2,(k)})$
\smallskip
\par
$\dot\psi^{2,(k)} = \Delta^{2,k}(\psi^{1,(k)},\psi^{2,(k)})$
\smallskip
\par
$\dot \phi^{i} = \Delta^{i} (\lambda^{i})$
\smallskip
\par
$\dot \lambda^{i}= 0$
\medskip
\par
In other words the eigenvalues of $T$ define a privileged
coordinates frame reducing to quadratures all its
automorphisms, i.e. all dynamics that leave it invariant. 
\par
Further, for the discrete part of the spectrum of~~ $T$~~ the 
sym\-plec\-tic form,~  
$
\omega_{0} = \sum_{i} f_{i} (\lambda^{i}) \delta \lambda^{i}
\wedge\delta\phi^{i}
$ 
~, can be introduced with respect to which the dynamics 
is a Hamiltonian one.
\par
In next section the mentioned geometrical structures 
will be exhibited for the Schr\"odinger equation.

\section{Can\-onical Co\-ordinates for the Schr\-\"odinger Equa\-tion}
Although in an infinite dimensional symplectic manifold a 
Darboux's chart, {\it a priori} does not exist, for the 
Schr\"odinger equation:

\be\label{LS}
i\hbar{\partial \psi\over \partial t}=-{ {\hbar}^2\over 2m}
\triangle \psi + U({\bf r})\psi,
\ee
natural 
canonical coordinates $p$ and $q$ can be introduced.
\par
We introduce the real and the imaginary part of the wave 
function $\psi$ :
\bea\label{CC}
\cases{
p({\bf r},t) = {\sl Im}\psi({\bf r},t)\cr 
q({\bf r},t) = {\sl Re}\psi({\bf r},t)\cr}\nonumber,
\eea
and in this way $L_2(Q, {\bf C})$ is considered as the cotangent 
bundle of $L_2(Q, {\bf R})$.
\par
In these new coordinates, equation ($\ref{LS}$) takes the form:
\begin{equation}
{d\over dt}\pmatrix{p\cr q} = {1\over {\hbar}}\pmatrix{0&-1\cr 1& 0}
\pmatrix{{\delta H_1 \over\delta p}\cr {\delta H_1 \over \delta q}}
\end{equation}
where $H_1$ is defined by:
\begin{equation}
H_1[q,p]:={1\over 2}\int
d{\bf r}\{{{\hbar}^2\over 2m}[(\nabla p)^2 + (\nabla q)^2]+U({\bf r})
(p^2+q^2)\}
\end{equation}
and
${\delta H \over \delta q},~~{\delta H \over\delta p}$ denote
the components of the gradient of $H[q,p]$ with respect to the
real $L_2$ scalar product.
\par
Our system is then a Hamiltonian dynamical system with respect to 
the Poisson bracket defined for any two functionals $F[q,p]$ and 
$G[q,p]$ by:
\be\label{CPB}
\Lambda_1(\delta F, \delta G):=\{F,G\}_1:= {1\over {\hbar}}\int
d{\bf r}({\delta F \over \delta q}\cdot {\delta G \over \delta p}-
{\delta F \over \delta p}\cdot {\delta G \over \delta q})
\ee
\par
What is less known is that the previous one is not the only possible 
Hamiltonian structure . 
As matter of fact the Schr\"odinger equation can also be written as:

\begin{equation}
{d\over dt}\pmatrix{p\cr q} = {1\over {\hbar}}\pmatrix{0&-{\cal H}\cr 
{\cal H}& 0}\pmatrix{{\delta H_0 \over\delta p}\cr {\delta H_0 \over 
\delta q}}
\end{equation}
where $H_0$ is defined by:

\begin{equation}
H_0[q,p]:={1\over 2}\int
d{\bf r}(p^2+q^2)
\end{equation}
and ${\cal H}$ is the Schr\"odinger operator:
\begin{equation}
{\cal H}:=-{{\hbar}^2\over 2m}\triangle + U({\bf r})
\end{equation}
\par
It is then again a Hamiltonian dynamical systems with a new 
Poisson bracket 
of any two functionals $F[q,p]$ and $G[q,p]$
given by:
\begin{equation}
\Lambda_0(\delta F, \delta G):=\{F,G\}_0:= \int
d{\bf r}({\delta F \over \delta q}\cdot {\cal H}{\delta G \over \delta p}-
{\delta F \over \delta p}\cdot {\cal H}{\delta G \over \delta q})
\end{equation}
So, with the same vector field, we have two choices:
\begin{itemize}
\item
A phase manifold with a universal symplectic structure:
\begin{equation}
\omega_1:= \hbar\int d{\bf r}(\delta p\wedge \delta q)
\end{equation}
and a Hamiltonian functional depending on the classical potential.
\item
A phase manifold with a symplectic structure determined by the classical 
potential

\begin{equation}
\omega_0:= \hbar\int d{\bf r}({\cal H}^{-1}\delta p\wedge  \delta q)
\end{equation}
and the universal Hamiltonian functional representing the quantum 
probability.
\end{itemize}
The two brackets satisfy the Jacobi Identity, as the associated $2$-forms 
are closed for they
do not depend on the point ($\psi\equiv (p,q)$) of the {\it phase space}.
\par
We have then the relation:
\begin{equation}
{\delta H_1 \over \delta u}= {\check T}{\delta H_0 \over \delta u}
\end{equation}
where:
\begin{equation}
 {\check T}:= \Lambda_1^{-1}\circ\Lambda_0=\pmatrix{{\cal H}&0\cr 0& {\cal H}}
\end{equation}
and
\begin{equation}
{\delta H \over \delta u}=
\pmatrix{{\delta H \over \delta q}\cr {\delta H \over \delta p}}
\end{equation}
As the tensor field T does not depend on the point ($\psi\equiv (p,q)$) 
of the phase space, its
torsion is identically zero, so that the relation (14) can be iterated to:
\begin{equation}
{\delta H_n \over \delta u}= {\check T}^n{\delta H_0 \over \delta u}
\end{equation}
It turns out that the Schr\"odinger equation admits infinitely many 
conserved functionals defined by:
\begin{equation}
H_n[q,p]:={1\over 2}\int 
d{\bf r}(p{\cal H}^np+q{\cal H}^nq)\equiv\int
d{\bf r}(\bar\psi{\cal H}^n\psi)
\end{equation}
They are all in involution with respect to the previous Poisson brackets:
\begin{equation}
\{H_n,H_m\}_0= \{H_n,H_m\}_1=0 
\end{equation}

It is worth to stress that for smooth potentials $U(x)$ in one space 
dimension, the eigenvalues of the Schr\"odinger operator ${\cal H}$  
are not degenerate and so the eigenvalues of $T$ are double degenerate.

\subsection{The eikonal transformation}
The transformation:

\be
\cases{
p({\bf r},t) = A({\bf r},t)sinS({\bf r},t)\hbar^{-1}\cr
q({\bf r},t) = A({\bf r},t)cosS({\bf r},t)\hbar^{-1}\cr}
\ee
between the $(p,q)$ coordinates and $(\pi= S(2\hbar)^{-1}J, \chi=A^2)$,
is a canonical transformation  as:
\begin{equation}
\delta p\wedge \delta q = \delta ({S\over 2\hbar})\wedge \delta A^2 
\end{equation}

The Hamiltonian $H_1$ becomes:
\begin{equation}
\tilde H_1[\chi, \pi]=\int d{\bf r}\{{\hbar^2\over 2m}({(\nabla\chi)^2
\over 4\chi}+4\chi(\nabla\pi)^2)+U\chi\}
\end{equation}
and Hamilton's equations:

\begin{equation}
\left\{
\begin{array}{l}
{\partial \pi\over \partial t}=-{1\over {\hbar}}{\delta\tilde H_1 \over 
\delta \chi}\\{\partial \chi\over \partial t}=~{1\over {\hbar}}
{\delta\tilde H_1 \over\delta \pi}
\end{array}\right.,
\end{equation}
give: 

\begin{equation}
\left\{
\begin{array}{l}
{\partial \pi\over \partial t}={\hbar\over 2m}{\triangle(\sqrt\chi) 
\over \sqrt\chi}-
{\hbar\over m}(\nabla\pi)^2-U\hbar^{-1}\\
{\partial \chi\over \partial t}=-{2\hbar\over m}div(\chi\nabla\pi) 
\end{array}
\right.
\end{equation}
where  $P=\chi$ and ${\bf J}=\hbar\chi{\nabla S\over m}$ represent the 
{\it probability density} and the {\it current density} respectively.
\par
This transformation being nonlinear will transform previous bi\-Hamil\-tonian
descriptions into a mutually compatible pair of nonlinear type. They are
of $C$-type as introduced by Calogero \cite{Ca91}.
\par
Finally, it is worth to stress that the Schr\"odinger equation, in  spite 
of its linearity, shows that the class of completely integrable field 
theories in higher dimensional spaces is not empty.

\section{The nonlinear Schr\"odinger equation}

The two-dimensional nonlinear Schr\"odinger equation:

\be\label{NLS}
i\hbar{\partial \psi\over \partial t}=- {{\hbar}^2\over 2m}
\psi_{xx} + b|\psi|^2\psi,
\ee
in the canonical  coordinates
\bea
\cases{
p(x,t) = {\sl Im}\psi(x,t)\cr 
q(x,t) = {\sl Re}\psi(x,t)\cr}\nonumber
\eea
takes the form:
\be
\pmatrix{p\cr q} = {1\over {\hbar}}\pmatrix{0&-1\cr 1& 0}
\pmatrix{{\delta K_1 \over\delta p}\cr {\delta K_1 \over \delta q}}
\end{equation}
where $K_1$ is defined by:
\begin{equation}
K_1[q,p]:={1\over 2}\int
dx\{{{\hbar}^2\over 2m}[(\partial_x p)^2 + b(\partial_x q)^2]+
(p^2+q^2)^2\}
\end{equation}

\par
It  is then a Hamiltonian dynamical system with respect to 
the canonical Poisson bracket $\Lambda_1$ defined in ($\ref{CPB}$):
\begin{equation}
\Lambda_1(\delta F, \delta G):=\{F,G\}_1:= {1\over {\hbar}}\int
dx({\delta F \over \delta q}\cdot {\delta G \over \delta p}-
{\delta F \over \delta p}\cdot {\delta G \over \delta q})
\end{equation}
\par
The previous one is not the only possible 
Hamiltonian structure . 
As matter of fact the nonlinear Schr\"odinger equation can also be written as:

\begin{equation}
{d\over dt}\pmatrix{q\cr p} = {\cal H}_N
\pmatrix{{\delta K_0 \over\delta q}\cr {\delta K_0 \over 
\delta p}}
\end{equation}
where ${\cal H}_N$ is the {\it Poisson operator}:
\be
{\cal H}_N={1\over {\hbar}}
\pmatrix{-{\hbar\over\sqrt{2m}}\partial_x+2\alpha pD^{-1}p&~-2\alpha pD^{-1}q\cr 
-2\alpha qD^{-1}p& -{\hbar\over\sqrt{2m}}\partial_x+2\alpha qD^{-1}q}
\ee
with~ $\alpha=b{\sqrt{2m}\over \hbar}$~, and 

\be
D^{-1}:= {1\over 2}(\int_{-\infty}^x-\int_x^\infty )~~~;~~~~~
K_0[q,p]:={\hbar\over\sqrt{2m}}\int dx(qp_x)
\ee

\par
It is then again a Hamil\-tonian dynamical system with a new 
Pois\-son bra\-cket~\footnote{For simplicity the proof that 
$\Lambda_2$ satisfy the Jacobi
Identity is omitted.} 
of any two functionals $F[q,p]$ and $G[q,p]$
given by:
\begin{equation}
\Lambda_2(\delta F, \delta G):=\{F,G\}_2:= \int
dx({\delta F \over \delta q}\cdot {\cal H}_N{\delta G \over \delta p}-
{\delta F \over \delta p}\cdot {\cal H}_N{\delta G \over \delta q})
\end{equation}

Once again, with the same vector field, we have two choices:
\begin{itemize}
\item
A phase manifold with the canonical symplectic structure:
\begin{equation}
\omega_1\equiv \Lambda_1^{-1}:= \hbar\int dx(\delta p\wedge \delta q)
\end{equation}
and a Hamiltonian functional accounting for the interaction.
\item
A phase manifold with a symplectic structure determined by the
interaction

\be
\omega_2\equiv \Lambda_2^{-1}:= \hbar\int dx({\cal H}_N^{-1}\delta p\wedge  \delta q)
\ee
and a {\it free} Hamiltonian functional given by the mean value
of the momentum $\hat p= -i\hbar\partial_x$.
\end{itemize}

We have then the relation:
\be\label{RS}
{\delta K_1 \over \delta u}= {\check T}_N{\delta K_0 \over \delta u}
\ee
where:
\be
 {\check T}:= \Lambda_1^{-1}\circ\Lambda_2=\pmatrix{2\alpha qD^{-1}p&{\hbar\over\sqrt{2m}}
\partial_x+2\alpha qD^{-1}q\cr 
-{\hbar\over\sqrt{2m}}\partial_x+2\alpha pD^{-1}p&-2\alpha pD^{-1}q}
\ee
\par
It can be shown that the sum $\Lambda_2+\Lambda_1$ is again 
a Poisson bracket. This is equivalent to the vanishing of  
the torsion of the tensor field $T_N$ , so that the relation 
($\ref{RS}$) can be iterated to:
\be\label{RR}
{\delta K_n \over \delta u}= {\check T_N}^n{\delta K_0 \over \delta u}
\ee
It turns out that the nonlinear Schr\"odinger equation admits infinitely many 
conserved functionals.
\medskip
\par
First three functionals are:
\bigskip

\bea
&&K_{-1}[q,p]:={1\over 2}\int 
dx(p^2+q^2)\equiv \int
dx(\bar\psi\psi)\cr
&&K_0[q,p]:=\int 
dx(qp_x)\equiv 2i\int
dx(\bar\psi\psi_x)\cr
&&K_1[q,p]:={1\over 2}\int
dx\{{{\hbar}^2\over 2m}[(\partial_x p)^2 + (\partial_x q)^2]+
(p^2+q^2)^2\}
\eea
\bigskip

They are all in involution with respect to the previous Poisson brackets:
\begin{equation}
\{K_n,K_m\}_0= \{K_n,K_m\}_1=0 
\end{equation}
\medskip

Observing that
\be
{\delta K_0 \over \delta u}= {\check T_N}{\delta K_{-1} \over \delta u}
\ee
the recursion relation (\ref{RR}) can be completed to:
\be
{\delta K_n \over \delta u}= {\check T_N}^{n+1}{\delta K_{-1} \over \delta u}
\ee

Turning back to the complex notation, with $\hbar=2m=\alpha=1$, we have 
the general sheme of the next page.
\vfill\eject

\begin{center}
{\sl Schr\"odinger Hierarchy}
\par
$\vdots$
\par
\fbox{\parbox{1.9 cm}{$\dot\psi=-i{\cal H}^2\psi$}}	
\par
$\uparrow$
\par
$\hat T$
\par
$|$
\par
\fbox{\parbox{1.8 cm}{$\dot\psi=-i{\cal H}\psi$}}		
\par
$\uparrow$
\par
$\hat T$
\par
$|$
\par
\fbox{\parbox{1.8 cm}{\fbox{\parbox{1.5 cm}{$\dot\psi=-i\psi$}}}}	
\par
$|$
\par
$\hat T_N$
\par
$\downarrow$
\par
{\sl s-Gordon Hierarchy}$\longleftarrow$
$\hat T_G$--
\fbox{\parbox{1.2 cm}{$\dot\psi=\psi_x$}}	
-- $\hat T_K$$\longrightarrow$
{\sl KdV Hierarchy~~~~~}
\par
$|$
\par
$\hat T_N$
\par
$\downarrow$
\par
\fbox{\parbox{3.5 cm}{$\dot\psi=i(\psi_{xx}+|\psi|^2\psi)$}}
\par
$|$
\par
$\hat T_N$
\par
$\downarrow$
\par
\fbox{\parbox{3.8 cm}{$\dot\psi=-(\psi_{xxx}+3|\psi|^2\psi_x)$}}
\par
$|$
\par
$\hat T_N$
\par
$\downarrow$
\par
\fbox{\parbox{10 cm}{$\dot\psi=-i(\psi_{xxxx}+4|\psi|^2\psi_{xx}+
3\bar\psi\psi_x^2+2\psi|\psi_x|^2+\psi^2\bar\psi_{xx}+{3\over 2}|\psi|^4\psi)$}}
\par
$\vdots$
\par
{\sl Nonlinear Schr\"odinger Hierarchy}
\end{center}
\bigskip
\par\noindent
$\hat T~~\bullet:=-\triangle\bullet+U\bullet\equiv {\cal H}\bullet$
\par\noindent
$\hat T_G\bullet:=\partial_{xx}\bullet+\psi_x D^{-1}\psi_x\bullet$
\par\noindent
$\hat T_K\bullet:=\partial_{xx}\bullet+{2\over 3}\psi\bullet + {1\over 3}\psi_x D^{-1}\bullet$
\par\noindent
$\hat T_N\bullet:=i(\partial_x\bullet+\psi D^{-1}[\psi\bar{(\bullet)}+\bar\psi (\bullet)])$
\newpage

\section{Concluding Remarks}
\par
It is interesting to observe that $K_{-1}$ is a conserved functional both
for the Schr\"odinger and the nonlinear Schr\"odinger equations. The
same is not true for $K_0$. This is due the fact that Schr\"odinger equation
is not invariant under space translations and  $K_0$ corresponds to
the mean value $<\hat p>$ of the linear momentum $\hat p=-ih\partial_x$.
In other words the vector field associated to $K_0$
{\it via} the canonical Poisson bracket $\Lambda_1$ is invariant for translation.

It is worth finally to compare the recursion operators 
of the Schr\"odinger, with vanishing potential $U(x)$, and Nonlinear 
Schr\"odinger, with $\alpha=0$,  hierarchies. It turns out that $\hat T=\hat T_N^2$.

\section*{Acknowledgments}
The author wishes to thank prof. M.Rasetti for his interest and comments,
Prof.s S.De Martino and S.De Siena, Dr.s R.Fedele and G.Miele for useful remarks.

\end{document}